\documentclass[reprint,prl,nolinenumbers,superscriptaddress]{revtex4-1}

\usepackage{hyperref}
\usepackage[english]{babel}
\usepackage[T1]{fontenc}
\usepackage[latin1]{inputenc}
\usepackage{amsfonts,amsmath,amssymb}
\usepackage{graphicx}
\usepackage{subfigure}
\usepackage{epsfig}
\def\imagetop#1{\vtop{\null\hbox{#1}}}

\begin{document}

\title{Supershear Rayleigh waves at a soft interface}
\author{Anne Le Goff}
\affiliation{Microfluidique, MEMs et Nanostructures, UMR Gulliver 7083, ESPCI, 75005 Paris, France}
\author{Pablo Cobelli}
\affiliation{Departamento de F\'\i sica, Facultad de Ciencias Exactas y Naturales, Universidad
de Buenos Aires and IFIBA, CONICET, Ciudad Universitaria, 1428 Buenos Aires, Argentina}
\affiliation{Physique et M\'ecanique des Milieux H\'et\'erog\`enes PMMH, UMR CNRS 7636-ESPCI-UMPC 
Univ. Paris 6--UPD Univ. Paris 7}
\author{Guillaume Lagubeau}
\affiliation{Physique et M\'ecanique des Milieux H\'et\'erog\`enes PMMH, UMR CNRS 7636-ESPCI-UMPC 
Univ. Paris 6--UPD Univ. Paris 7}
\affiliation{Departamento de F\'\i sica, Universidad de Santiago de Chile, Santiago de Chile}
\date{\today}

\begin{abstract}
We report on the experimental observation of waves at a liquid foam surface propagating faster than the bulk shear waves. The existence of such waves has long been debated, but the recent observation of supershear events in a geophysical context has inspired us to search for their existence in a model viscoelastic system. An optimized fast profilometry technique allowed us to observe on a liquid foam surface the waves triggered by the impact of a projectile. At high impact velocity, we show that the expected subshear Rayleigh waves are accompanied by faster surface waves that can be identified as supershear Rayleigh waves.
\end{abstract}

\maketitle

Supershear rupture, i.e., rupture propagating at unusual speed, faster than that of shear waves, has recently been observed in large seismic events \cite{Bouchon03} and experiments \cite{Xia04}. These events break the theoretical limit stating cracks should not propagate faster than the velocity of Rayleigh waves (transverse surface waves) \cite{Freund98}. Their increased hazard raised `the need to study speed' \cite{Das07} and they were successfully explained using both internal friction and soils heterogeneities. Inspired by these remarkable observations, we aim to address a related question: could Rayleigh waves \cite{Rayleigh1885} themselves propagate at a velocity faster than the shear waves? The existence of supershear Rayleigh waves has long been debated, either in pure elastic \cite{Rayleigh1885,Gilbert62,Hayes62,Hudson80,Schroder01} or dissipative media \cite{Scholte47,Currie77,Romeo02}, but lacks experimental observations.

\begin{figure}
\centering
\includegraphics[scale=0.35,trim=65 230 0 0,clip]{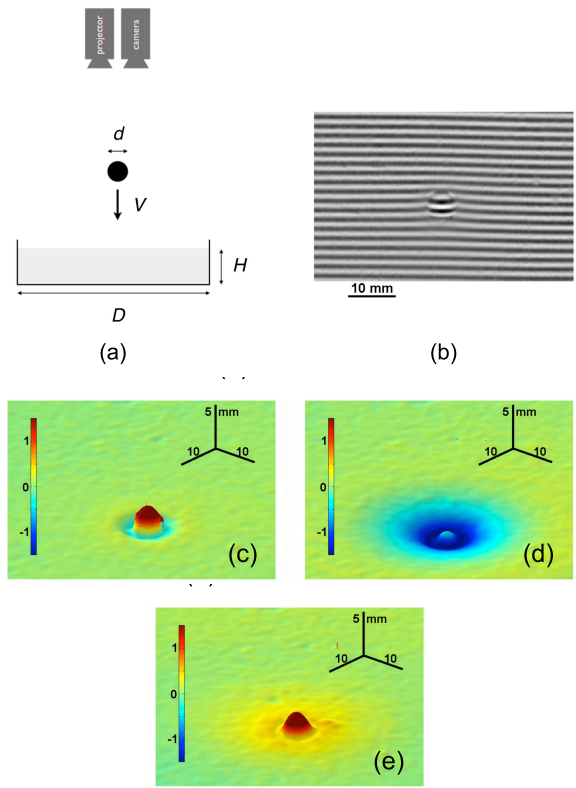}\\
\includegraphics[scale=0.4,trim=2 50 528 395,clip]{Figure1}
\includegraphics[scale=0.4,trim=265 50 265 395,clip]{Figure1}\\
\includegraphics[scale=0.4,trim=528 40 0 395,clip]{Figure1}
\caption{Map of the foam's vertical displacement. (a) Sketch of the experimental setup 
($0.8 < V < 6.5$~m/s, $d = 6$~mm, $H = 65$~mm and $D = 115$~mm). (b) Example of deformed fringe pattern (corresponding height map shown in  (d)). Panels (c), (d) and (e) show foam surface height maps after impact for $V = 1.82$~m/s at times $t = 2.3$,~8.9 and 23.5~ms, respectively.}
\label{figure1}
\end{figure}

The model viscoelastic material that we use to address experimentally the aforementioned question is a liquid foam \cite{Hohler05}. It consists of a concentrated dispersion of gas bubbles in a continuous liquid phase. With its low shear modulus, causing waves to propagate slowly, foam is ideal for wave velocity measurements. Furthermore, because of its low density, inertia is negligible. However, strong dissipation makes acoustics methods challenging in viscoelastic media such as foams \cite{Mujica02}, and, although the rheology of these materials is widely studied \cite{Hohler05,Gopal99}, little is known about the propagation of surface waves in foams. Using the impact of a solid bead, we trigger deformations of a foam surface, that we measure by a fringe projection profilometry technique \cite{Cobelli09,maurel2009experimental} already tested for fast impacts on liquids \cite{lagubeau2010flower} as shown in Fig.~\ref{figure1}(a,b). We demonstrate that this non-invasive technique can accurately track impact-generated surface waves. For gentle impacts, our measurements are in agreement with the speed predicted from published rheology experiments \cite{Krishian10}. But the foam response changes as impact velocity increases, and some surface waves are found to propagate faster than shear waves. Finally, supershear Rayleigh waves may very well propagate at the surface of soft media. 

\begin{figure*}[ht!]
\includegraphics[scale=0.4,trim=0 50 428 0,clip]{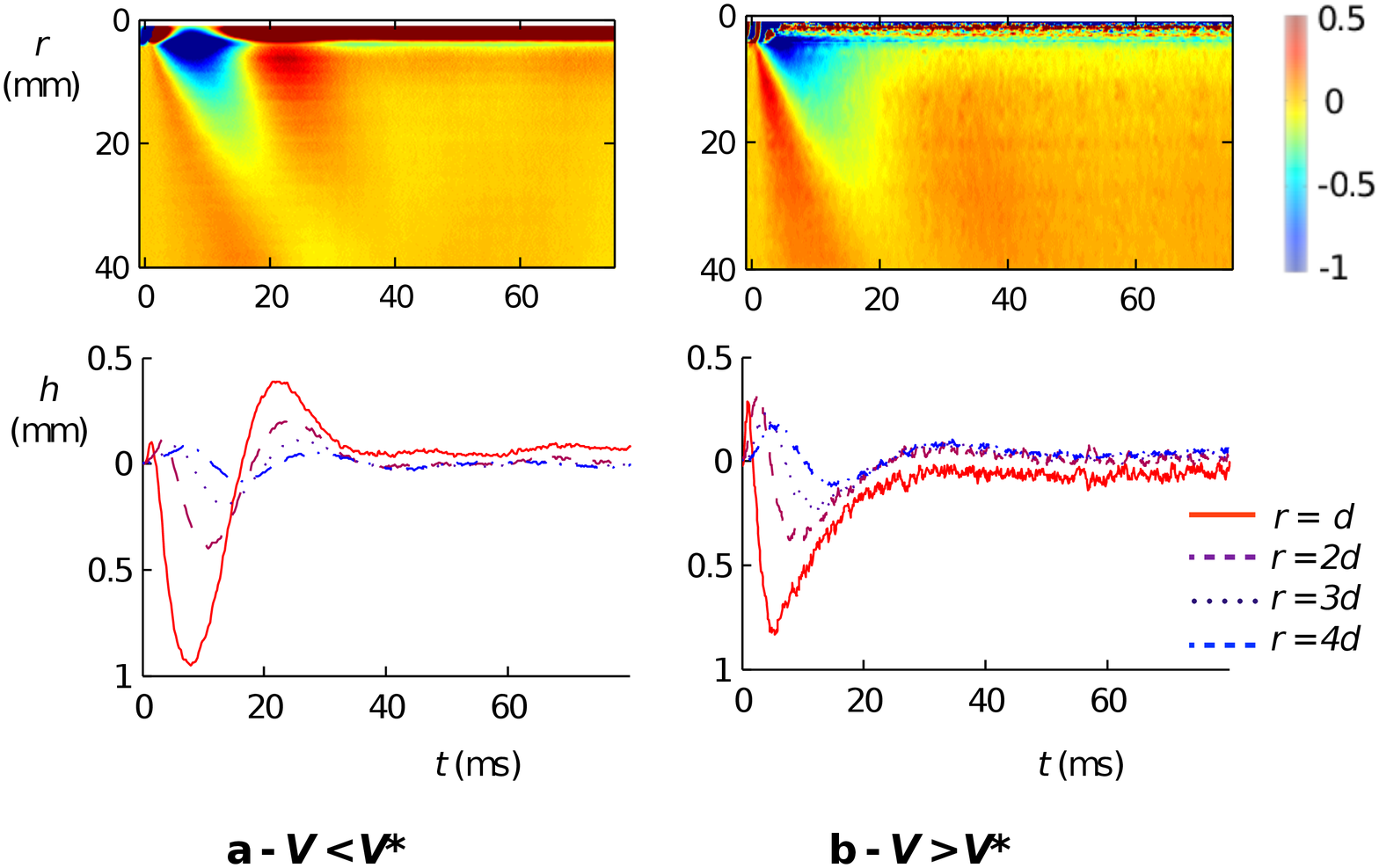}\hspace{10mm}
\includegraphics[scale=0.4,trim=400 50 0 0,clip]{Figure2}\\
{\small (a) $V < V^{*}$ \hspace{5cm} (b) $V > V^{*}$ }
\caption{Spatiotemporal analysis of surface height in two cases:  (a) $V  < V^*$ ($V = 1.82$~m/s). 
 (b) $V > V^*$ ($V = 5.53$~m/s). Top: spatiotemporal diagrams of $h(r, t)$, with time $t$ in horizontal direction and distance to impact center $r$ in vertical direction. Bottom: $h(r, t)$ at various distances~$r$, gradually increasing from one sphere diameter $d$ (red curve) to $4d$ (blue curve).}
\label{figure2}
\end{figure*}

\begin{figure}
\begin{tabular}{l l}
    \imagetop{\includegraphics[scale=1.15,trim=0 20 165 0,clip]{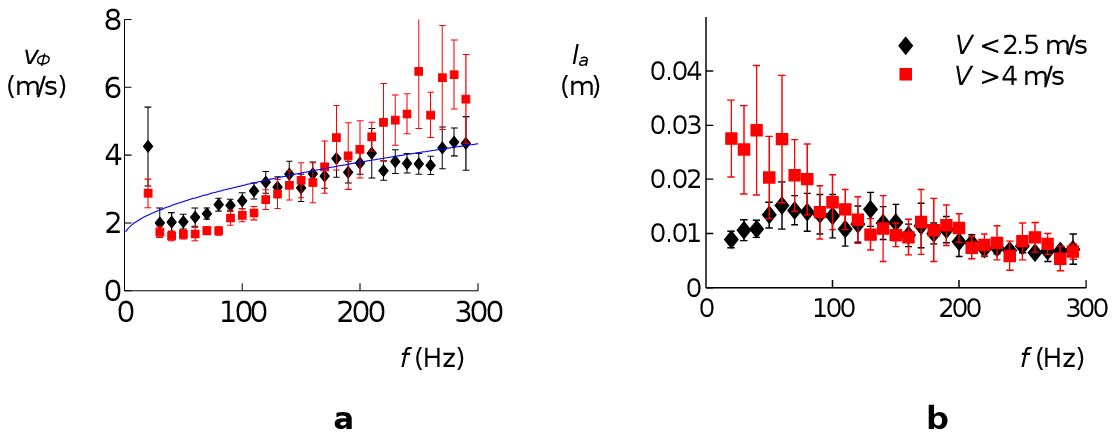}} & \imagetop{(a)} \\
    \imagetop{\includegraphics[scale=1.15,trim=160 20 0 0,clip]{Figure3}} & \imagetop{(b)}
\end{tabular}
\caption{Complex dispersion relation components.  (a) Phase velocity.  (b) Attenuation length. Black lozenges: slow impact case. Red squares: fast impact case. The error bars correspond to statistics over 12 experiments of distinct impact speeds and foam realizations. Blue line: predictions of shear wave speed $V_{S}$ based on published law for Gillette foam's shear modulus \cite{Krishian10}.}
\label{figure3}
\end{figure}

\textit{Materials and methods}.
A plastic sphere of diameter $d=6$~mm and density 1.2~g/cm$^3$ impacts normally the surface of a liquid foam, whose instantaneous vertical deformations are measured by means of a space-time resolved profilometry technique, as shown in Fig.~\ref{figure1}(a) and (b). We use the well-characterized ``Foamy Regular'' cream from Gillette 
\cite{Gopal99,Mujica02,Krishian10,Durian91}. The dimensions of the cylindrical container avoid side effects: its diameter $D = 11.5$~cm is much larger than the typical observed attenuation length $l_a$ (about 1~cm - see Fig.~\ref{figure3} (b)) and its depth $H = 6.5$~cm is larger than wavelength $\lambda$ as soon as $f > 30$~Hz (`deep foam' region). After filling the tank, a constant waiting time of 30~minutes is observed for the foam to reach a reproducible density of 77~Kg/m$^{3}$. The profilometry apparatus consists in a high resolution video projector (Epson TW5500), projecting a one dimensional fringe pattern on the foam surface, and a fast camera (Phantom v9.0) recording its deformations at 5000~fps. Using the parallel optical configuration as in Fourier Transform Profilometry technique, the displacement map can be deduced from the local phase changes of the fringe pattern \cite{Takeda83} knowing the optical axes separation (23.0~cm), the entrance pupil height (98.5~cm) and the projected fringe wavelength (1.82~mm). For better spatial accuracy, the phase extraction algorithm used Empirical Mode Decomposition filtering \cite{Huang98} instead of the usual Fourier filter \cite{Takeda83}. The temporal resolution is 0.2~$\mu$s The typical height resolution is 75~$\mu$m (0.05~rad in phase) at each point of a grid of 225~$\mu$m mesh size, and decrease to 10~$\mu$m after angular averaging. Examples of instantaneous height maps are shown in Fig.~\ref{figure1} and in supplementary video 2.

\textit{Results}. Figures~\ref{figure1}(c-e) display experimental height maps of the sample surface i.e., both the foam and the top of the sphere, for an impact velocity of $V = 1.82$~m/s. Shortly after impact, a small bump (about 100~$\mu$m amplitude) propagates radially (Fig.~\ref{figure1}(c), $t = 2.3$~ms). The sphere entering the foam then creates a dynamic crater, visible on Fig.~\ref{figure1}(d) ($t = 8.9$~ms), that disappears on Fig.~\ref{figure1}(e) ($t = 23.5$~ms) after the sphere "bounces" and its top becomes visible again above the foam surface. One can immediately notice that this deformation is axisymmetric around the impact point. We average measurements over all orientations to obtain $h(r,t)$ as shown in Fig.~\ref{figure2}(a) in a spatiotemporal diagram of this very experiment. The amplitude of deformations in this particular case is about 1~mm and remains well below $d$ in all our experiments, provided that observation is made far enough from the impact point, at a distance $r > 0.75 d$. For clarity, in the bottom image, we isolate 4~distances ($d, 2d, 3d$ and $4d$), for which we plot the time evolution $h(t)$. For $r = d$, the signal $h(t)$ looks like a strongly attenuated oscillatory motion, with a period of approximately 30~ms. At larger distances the signal has a similar shape, but delayed and damped. These four curves comparison provides evidence for damped propagative surface waves with a typical wave speed of a few meters per second and centimetric attenuation length. At an higher velocity of 5.53~m/s, the shape of $h(t)$ is qualitatively different, as illustrated in Fig.~\ref{figure2}(b). The second maximum present in Fig.~\ref{figure2}(a) disappears and the long time dynamics is then similar to that of an underdamped oscillator. We notice that, surprisingly, while the impact velocity is multiplied by 3 and therefore the kinetic energy by 9, the crater at $r = d$ remains of comparable depth.

Using wave analysis tools, we now characterize each of these two regimes. 
Adjusting the curves of $\hat{h}_{f}(r)$ and $\Phi_{f}(r)$, the amplitude and phase of the temporal
Fourier transform of $h(r,t)$, we can measure the phase velocity $V_{\phi}(f)$ 
and attenuation length $l_{a}(f)$ defined by 

\begin{equation*}
V_{\phi}(f) = \frac{2\pi f}{d\Phi_{f}/dr} \quad \text{and} \quad \hat{h}_{f}(r) \sim \frac{1}{\sqrt{r}}
\: \text{e}^{-{r}/{l_{a}}}.
\end{equation*}%

\begin{figure}
\begin{tabular}{l l}
    \imagetop{\includegraphics[scale=0.4,trim=0 30 425 0,clip]{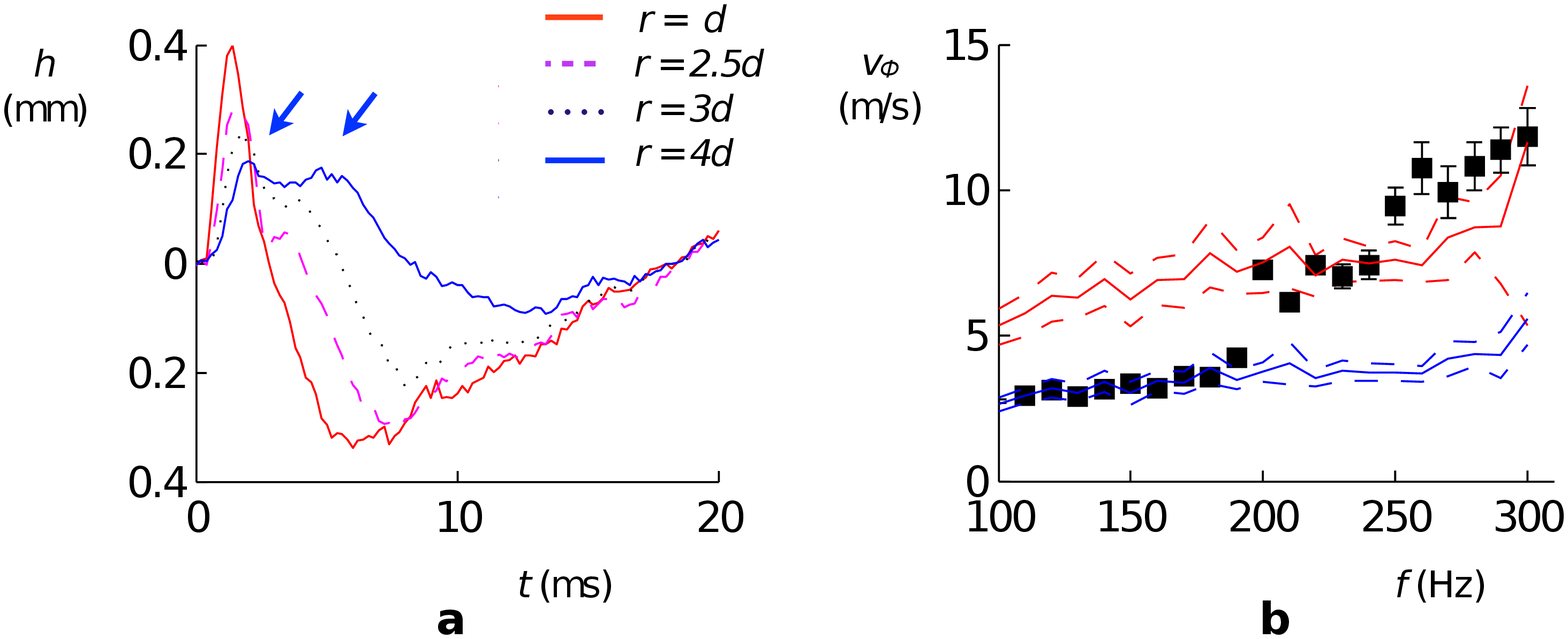}} & \imagetop{(a)} \\
    \imagetop{\includegraphics[scale=0.4,trim=425 33 0 0,clip]{Figure4}} & \imagetop{(b)}
\end{tabular}
\caption{Evidence of two different waves after a very fast impact (5 mm diameter bead, $V = 18.3$~m/s).  (a) Detail of $h(r, t)$ at early times for values of $r$ ranging from $2d$ to $4d$. (b) Black squares: phase velocity dispersion relation (Error bars corresponding to the confidence interval of the linear regression of $\phi(r)$). Blue continuous line: dispersion relation of the anelastic Rayleigh waves. Red continuous line: Prediction of the supershear wave velocity from the slow impact dispersion relation, using (\ref{eq:VSR}). The confidence interval (dashed lines) is defined as in Fig.~\ref{figure3}.}
\label{figure4}
\end{figure}

Both $V_{\phi}(f)$ and $l_{a}(f)$ dispersion curves are shown in Fig.~\ref{figure3}. Because amplitude decreases with frequency, the recorded deformations become smaller than our detection threshold for $f >300$~Hz. 
The complex formalism allows us to simultaneously account for the elastic and dissipative effects by defining the complex wave number $k_{c}(f)=V_{phi}(f)/2\pi f - i/l_{a}(f)$. 
The wavelength $\lambda=2\pi/\Re(k_{c})$ and attenuation length $l_{a}$ are found to be of about~1~cm, much larger than the bubble size (22~$\mu$m) and smaller than the tank dimensions. 
Therefore, we consider the foam as a semi-infinite homogeneous viscoelastic medium, whose storage and loss moduli may depend on frequency. This assumption is further justified by the fact that resonant effects resulting from the discrete structure of the foam only appear at frequencies of about $40\mbox{ kHz}$, much higher than our observation range \cite{BenSalemSM2013}. At the surface of such a material, the velocity of Rayleigh waves $V_{R}$ is a function of Poisson modulus $\nu$ and bulk shear wave velocity $V_{S}$, and is always smaller than $V_{S}$ \cite{Carcione01}. In the case of an incompressible material such as the foam used in this experiment, $\nu = 0.5$ and $V_{R}(f) \sim 0.955 \: V_{S}(f)$ \cite{Rayleigh1885}. In the slow impact case, $V_{\phi}$ increases from 2~m/s at 40~Hz to 4~m/s at 300~Hz. The shape of $V_{\phi}(f)$ curve is also found independent on the sphere's diameter. The blue line in Fig.~\ref{figure3} represents a prediction of the velocity $V_{S}(f) = \sqrt{\mu(f)/\rho}$ of bulk shear waves based on a standard viscoelastic model \cite{PhysRevLett.76.3017} and published parameters for Gillette foam's shear modulus $\mu$ \cite{Krishian10}. The fact that $V_{\phi}(f)$ is very close to $V_{S}(f)$ is consistent with the expected observation of Rayleigh waves. Dissipative counterparts of the Rayleigh waves propagating at solid elastic surfaces are called quasi-elastic Rayleigh waves \cite{Carcione01}. In the fast impact case, we detect a significantly higher phase velocity when $f > 190$~Hz. Surprisingly, some waves travel much faster than bulk shear waves. The reason of this increase is ambiguous in the framework of monomodal Fourier analysis: either a non-linear increase of elastic modulus modifies the speed of shear and Rayleigh waves, or the superposition of a second, faster, wave creates a larger apparent velocity. By further increasing impact velocity, we obtain the $h(t)$ curves shown in Fig.~\ref{figure4} (a) where two bumps propagating at different velocities are clearly distinguishable. This allows us to eliminate the first option. The slower bump speed is comparable to the quasi-elastic Rayleigh wave speed, while the other bump travels roughly twice as fast. For $r < 3d$, the fast bump predominates, and the monomodal analysis can be applied for measuring its dispersion relation (Fig.~\ref{figure4} (b)). If $f < 200$~Hz, it again coincides with the quasi-elastic Rayleigh wave dispersion relation but for larger frequencies $V_{\phi}$ jumps to nearly twice the expected velocity.

\textit{Discussion}.
We now discuss the nature of these fast surface waves. A liquid-like wave is first considered. Soft interfaces can mix elastic and liquid surface properties as in agarose gel, on which a pseudo-elastic wave and a pseudo-capillary wave can coexist \cite{Monroy98}. For comparing the elastic-like and liquid-like contribution we define the ratio of the surface stress induced by elastic surface waves and liquid surface waves \cite{Ahn02}: 
$\beta = (\gamma k + \rho g/k)/\mu$ with $\mu$ the shear modulus, $\gamma$ the surface tension and the real wave number $k = \Re(k_{c})$. For the agarose experiments \cite{Monroy98,Ahn02} $\beta>1$ while for liquid foam we find $\beta < 0.02$. We conclude that liquid-like surface waves do not match our observations. A second possible nature is a different kind of elastic surface wave: Rayleigh waves are the real solution of a secular equation, for which a complex solution has also been proposed, for elastic (`leaky waves') \cite{Gilbert62,Hudson80} or viscoelastic (`viscoelastic waves') half-spaces \cite{Scholte47,Currie77,Carcione01}. For unity, we name complex solutions `supershear Rayleigh waves'. They correspond to waves propagating faster than the shear waves and weakly transverse in polarization. They may exist only in a limited range of Poisson modulus $\nu$ and shear wave quality factor, defined as $Q=-\Re(k_{c}^{2})/\Im(k_{c}^{2})$: 
$\nu > 0.26$ in any case \cite{Schroder01,Carcione01} and $Q < 6.29$ for the viscoelastic case \cite{Carcione01}. However, the physical admissibility of this solution has been debated \cite{Hayes62,Currie77,Lai02}. It has been rigorously proven spurious for plane waves but a careful examination of the radiation condition for a source localized in time and space (as for geological events or impacts) showed that it can then be acceptable \cite{Schroder01}, and a few fast waves propagating on soils have also been reported \cite{Roth99,Martin06}. With $\nu = 0.5$ and $Q \approx 1.32$, our experiments are well inside the theoretically expected existence domain of supershear Rayleigh waves. To go further, we predict the supershear Rayleigh wave velocity, starting with the measurement of $V_{R}$ in the slow impact case. If $V_{SR}$ is the complex wave speed of the supershear Rayleigh wave, then for an incompressible medium, 
$V_{R}^2 \approx 0.91 \: V_{S}^2$ and $V_{SR}^2 \approx (3.54 + 2.23 i) \: V_{S}^{2}$ \cite{Carcione01}. Eliminating $V_{S}$ leads to
\begin{equation}
V_{SR}^2 \approx (3.88 + 2.44 i) \: V_{R}^{2}.
\label{eq:VSR}
\end{equation}

Measurements of $V_R$ and subsequent predictions of $V_{SR}$ are plotted in Fig.~\ref{figure4}(b) (blue and red continuous lines). While the phase velocity after the 18.3~m/s impact is still close to $V_{R}$ for $f < 200$~Hz, it jumps to values in good agreement with the supershear wave speed $V_{SR}$ for higher frequencies. This is strong evidence that we observed the propagation of the supershear Rayleigh waves on a foam surface. They were visible for impact velocities faster than the measured shear wave velocities and only for $f > 200$~Hz. This might be due to a deep tank condition (experimentally $\lambda/H < 0.57 \pm 0.07$) related to the need of a companion bulk shear wave \cite{Schroder01}.

We conclude that supershear Rayleigh waves have been observed in our experiment. This implies that they might also be expected to propagate on other materials of similar $Q$. In particular, some water or oil filled soils \cite{Korneev04} of high seismic hazard \cite{Flores07}, or frictionally held soils. The existence of supershear waves had also been predicted, in the purely elastic case, i.e. without any restriction of $Q$ \cite{Schroder01}. If so, this would imply that these waves may be relevant for any type of high Poisson ratio soil, not only in soft dissipative ones.

\begin{acknowledgments}
 The authors wish to thank Agn\`es Maurel, Philippe Petitjeans and Vincent Pagneux for giving them access to the profilometry setup, Andr\'ea Laug\`ere and Babacar Gomis for their contribution to the preliminary experiments, Marie Le Merrer, Pablo Mininni and Meredith Root-Bernstein for careful reading of the manuscript, as well as Etienne Couturier, Paul Martin and members of the GDR Mousses for inspiring discussions.
\end{acknowledgments}

\bibliography{bib}

\bibliographystyle{apsrev4-1}

\end{document}